\documentclass[twocolumn,aps,pra]{revtex4}
\usepackage{epsfig}
\usepackage[english]{babel}
\usepackage{latexsym}
\usepackage{graphics}
\usepackage{subfigure}
\usepackage{graphicx}
\usepackage{dcolumn}
\usepackage{amsmath}
\usepackage{hyperref}
\usepackage{amssymb}
\usepackage{color}

\begin{document}

\title{Single-photon ionization of aligned H$_2^+$ with lower photon energy}

\author{F. B. Zhang$^{1,2,\ddag}$, J. Y. Che$^{2,\ddag}$,  W. Y. Li$^{1,*}$, C. Chen$^{2}$, and Y. J. Chen$^{2,\dag}$}

\date{\today}

\begin{abstract}

We study single-photon ionization of aligned H$_2^+$ in a high-frequency low-intensity laser field. We focus on the case where the laser frequency is not far larger than the ionization potential of the target. The calculated photoelectron momentum distribution through numerical solution of time-dependent Schr\"{o}dinger equation shows clear interference patterns. By developing a theory model applicable for high-frequency laser field, we show that the interference patterns can not be explained by the interference of the electronic wave with the observed momentum between these two atomic centers of the molecule. The Coulomb potential influences remarkably on the momentum of the emitting electronic wave responsible for this interference. Our results suggest a manner for probing the structure and the electron dynamics of the molecule in single-photon ionization.

\end{abstract}

\affiliation{1.School of Mathematics and Science and Hebei Key Laboratory of Photoelectronic Information and Goe-detection Technology, Hebei GEO University, Shijiazhuang, China\\
2.College of Physics and Information Technology, Shaan'xi Normal University, Xi'an, China}
\maketitle

\section{Introduction}
The interaction of atoms and molecules with a strong low-frequency laser field induces rich physical phenomena such as above-threshold ionization (ATI) \cite{Agostini1979, Yang1993, Paulus1994, Lewenstein1995, Becker2002}, non-sequential double ionization (NSDI) \cite{Niikura2003, Zeidler2005, Becker2012}, and high-order harmonic generation (HHG) \cite{McPherson1987, Huillier1991, Corkum1993, Lewenstein1994, Krausz2009}, etc.. The core step of these processes is tunneling ionization \cite{Keldysh,Faisal,Reiss}. When the Coulomb potential is bent by the intense laser field, the bound electron can escape from the laser-Coulomb-formed barrier through tunneling. Then the motion of the ionized electron can be described classically, as shown by the well-known semiclassical \cite{Corkum1993} or quantum \cite{Lewenstein1994} strong-field models. During the tunneling process characterized by an imaginary electronic momentum, the symmetry of the Coulomb potential is destroyed by the laser field and the structure of the target atom or the target molecule plays a relatively small role in shaping the continuum wave packet \cite{Chen2010,Itatani}. As a result, the structural information of the target can not be easily deduced from photoelectron momentum distribution (PMD) of ATI. Instead, the recombination process of the HHG encodes rich structural information of the target and can be used to read the information of the highest occupied electronic orbital through the procedure of HHG-based orbital tomography \cite{Itatani,Haessler2010,Vozzi2011,chen2013}.

Different from the case of the high-intensity and low-frequency laser field where the mechanism of tunneling ionization dominates,
the ionization of atoms or molecules in a low-intensity and high-frequency laser field is dominated by single-photon ionization.
For a diatomic molecule with two atomic centers, such as N$_2$ \cite{Rolles2005,Liu2006,Schoffler20081,Zimmermann2008},
H$_2$ \cite{Akoury2007,Schoffler2008,Canton2011,Waitz2016},
O$_2$ \cite{Liu2015}, and Ne$_2$ \cite{Sann2016,Kunitski2019},
during the process of single-photon ionization, the emitting electronic wave with real momentum interferes
between these two atomic centers of the molecule.
This phenomenon can be understood as the molecular-level double-slit interference. The resulting interference patterns in PMD encode the structural information of the molecule, such as the internuclear distance, the orbital symmetry and the alignment of the molecule. Most of such experiments are performed for high photon energy far larger than the ionization potential $I_p$ of the target molecule. For relatively small photon frequency, relevant single-photon ionization phenomena are less studied. One can expect that such studies can provide not only the ground-state information but also the continuum-state information of the molecule near the nuclei.

\section{Numerical methods}

In this paper, we  study single-photon ionization of aligned H$_2^+$ in a high-frequency and low-intensity laser field.  We first simulate the relevant ionization dynamics through numerical solution of the time-dependent Schr\"{o}dinger equation (TDSE).

In the length gauge, the Hamiltonian of the model H$_2^+$ system interacting with a laser field can be written as (in atomic units of $\hbar=e=m_e=1$) \begin{equation}
H(t)=H_0+\textbf{r}\cdot\textbf{E}(t).
\end{equation}
Here, the term $H_0=\textbf{p}^2/2+V(\textbf{r})$ is the field-free Hamiltonian and $\text{V}(\textbf{r})$ is the Coulomb potential of the molecule.
The potential used here has the form of
$V(\mathbf{r})=-Z/\sqrt{\zeta+\mathbf{r}_1^{2}}-Z/\sqrt{\zeta+\mathbf{r}_2^{2}}$
with $\mathbf{r}_{1(2)}^2=(x\pm \frac{R}{2}\cos\theta')^2+(y\pm \frac{R}{2}\sin\theta')^2$.
Here, $\zeta=0.5$ is the smoothing parameter which is used to avoid the Coulomb singularity, and $\theta'$ is the alignment angle.
$Z$ is the effective nuclear charge which is adjusted such that
the ionization energy of the model molecule at the internuclear distance $R$
is $I_p$=$1.1$ a.u.. Typically, for the equilibrium separation of $R=2$ a.u., we have $Z=1$.
The geometry of the model molecular system is presented in Fig. 1.

\begin{figure}[t]
\begin{center}
\rotatebox{0}{\resizebox *{6.5cm}{6cm} {\includegraphics {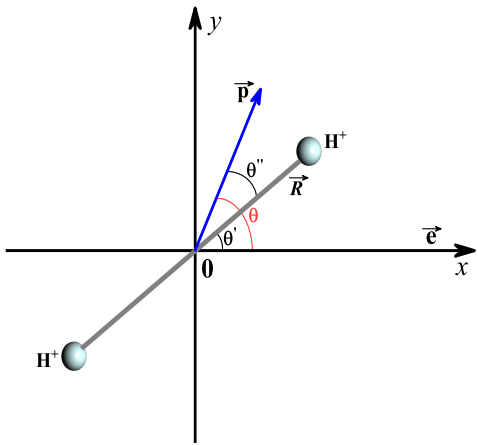}}}
\end{center}
\caption{A sketch of the molecular geometry and the coordinate system used in simulations.
The molecular axis of H$_{2}^{+}$ with the internuclear distance ${R}$ is located in
the $xoy$ plane with an angle $\theta'$ to the $x$ axis.
The center of mass of the molecular system coincides with the origin of the coordinate system.
The laser polarization $\textbf{e}$ is along the $x$ axis.
The photoelectron momentum $\textbf{p}$ has an angle $\theta$ to the laser polarization and an angle $\theta''$ to the molecular axis.
}
\label{fig:g1}
\end{figure}

The term $\textbf{E}(t)$ in Eq. (1) is the external electric field, which has the form of $\textbf{E}(t)=\textbf{e}E(t)$, with $E(t)=f(t)E_0\sin(\omega t)$. The unit vector $\mathbf{e}$ along the laser polarization is parallel to the $x$ axis, $E_0$ is the maximal laser amplitude related to the peak intensity $I$, $f(t)$ is the envelope function, and  $\omega$ is the laser frequency. 
We use trapezoid-shaped laser pulses with a total duration 400 optical cycles and linear ramps of 50 optical cycles.
The use of the multi-cycle long pulses here allows a high resolution of the photon frequency $\omega$ in our simulations.
The space step used here is $\Delta x=\Delta y=0.4$ a.u. with the grid size of $L_x=L_y=819.2$ a.u. and the time step used here is $\Delta t=0.05$ a.u..  A mask function $\cos^{1/8}$ is used for $|x|\geq300$ a.u. and
$|y|\geq300$ a.u. to absorb the continuum wavepacket at the boundary and obtain the PMD. More details for solving the TDSE of $i\dot{\psi}(\mathbf{r},t)=H(t)\psi(\mathbf{r},t)$ with spectral method \cite{Feit1982} and obtaining PMD can be found in \cite{Gao2017, Wang2017}.

\section{Analytical methods}
To analytically study single-photon ionization of a molecular system in a laser pulse with high laser frequency $\omega$ and low laser intensity $I$, we use the fundamental idea in strong-field approximation (SFA) \cite{Lewenstein1994,Lewenstein1995}. The SFA is based on three important assumptions. 1) Except the ground state of the system, the contributions of other bound states to the evolution of the system can be neglected. 2) The depletion of the ground state can be neglected. 3) The continuum state can be approximated with the plane wave.  Although the SFA is obtained for cases of high laser intensity and low laser frequency, the above assumptions can also be used for cases where 1) the laser frequency $\omega$ is larger than the ionization potential $I_p$ (i.e., $\omega>I_p$) so that  single-photon ionization is guaranteed and 2) the laser intensity is so low that the ionization is weak.
These two conditions are fulfilled in our cases.

In the SFA, the amplitude of the photoelectron with the drift momentum $\textbf{p} $ can be written as \cite{Lewenstein1995}
\begin{equation}
c(\textbf{p})=-i\int^{T_{p}}_0dt^\prime{\textbf{E}}(t^\prime)\cdot{\textbf{d}_i}{[\textbf{p}+\textbf{A}(t^\prime)]}e^{iS(\textbf{p},t^\prime)}.
\end{equation}
Here, the term  $S(\textbf{p},t')=\int_{}^{t'}\{{[\textbf{p}+\textbf{A}(t''})]^2/2+I_p\}dt''$ is the semiclassical action and $T_{p}$ is the length of the total pulse.
The term $\textbf{d}_i(\textbf{v})=\langle{\textbf{v}}|\textbf{r}\vert{{0}\rangle}$ denotes the dipole matrix element for the bound-free transition.
The term $\textbf{A}(t)=-\int^{t}\textbf{E}(t')dt'$ is the vector potential of the electric field $\textbf{E}(t)$. 

For cases of high-frequency laser field with $\omega>I_p$, considering $E(t)=E_0\sin\omega t=E_0(e^{i\omega t}-e^{-i\omega t})/(2i)$, we have $c(\textbf{p})=[c_2(\textbf{p})-c_1(\textbf{p})]/(2i)$ with
\begin{equation}
c_1(\textbf{p})=-i\int^{T_{p}}_0dt^\prime E_0 \textbf{e}\cdot{\textbf{d}_i}{[\textbf{p}+\textbf{A}(t^\prime)]}e^{iS_1(\textbf{p},t^\prime)}
\end{equation}
and
\begin{equation}
c_2(\textbf{p})=-i\int^{T_{p}}_0dt^\prime E_0 \textbf{e}\cdot{\textbf{d}_i}{[\textbf{p}+\textbf{A}(t^\prime)]}e^{iS_2(\textbf{p},t^\prime)}.
\end{equation}
Here,  $S_1(\textbf{p},t')=\int_{}^{t'}\{{[\textbf{p}+\textbf{A}(t''})]^2/2+I_p\}dt''-\omega t'$ and  $S_2(\textbf{p},t')=\int_{}^{t'}\{{[\textbf{p}+\textbf{A}(t''})]^2/2+I_p\}dt''+\omega t'$. 
The term $c_1(\textbf{p})$ satisfies the energy conservation
and therefore contributes mainly to ionization in the present high-frequency cases. The temporal integral in the expression of $c_1(\textbf{p})$ can also be evaluated by the saddle-point method \cite{Lewenstein1995,Becker2002},
with solving the following equation
\begin{equation}
[\textbf{p}+\textbf{A}(t_s)]^2/2=\omega-I_p.
\end{equation}
The solution $t_s=t_0$ of the above equation is real for $\omega>I_p$. 
We consider that the solution $t_0$ is also the saddle-point solution of Eq. (2).
The corresponding amplitude of the solution ($\textbf{p},t_0$) can be written as
\begin{equation}
F(\mathbf{p},t_0)\propto\big[\beta\textbf{E}(t_0)\cdot \textbf{d}_i(\textbf{p}+\textbf{A}(t_0))e^{iS(\textbf{p},t_0)}\big],
\end{equation}
with  $\beta\equiv({1/det(t_0)})^{1/2}$.
The term $det(t_0)$ in the definition of $\beta$ is the determinant of the matrix formed by the second derivatives of the action \cite{Lewenstein1995}.
The whole amplitude for photoelectron with a momentum $\textbf{p}$ can be written as
\begin{equation}
c(\textbf{p})\propto \sum_s F(\textbf{p},t_s).
\end{equation}
The sum runs over all possible saddle points $t_s=t_0$.

\begin{figure}[t]
\begin{center}
\rotatebox{0}{\resizebox *{8.5cm}{8cm} {\includegraphics {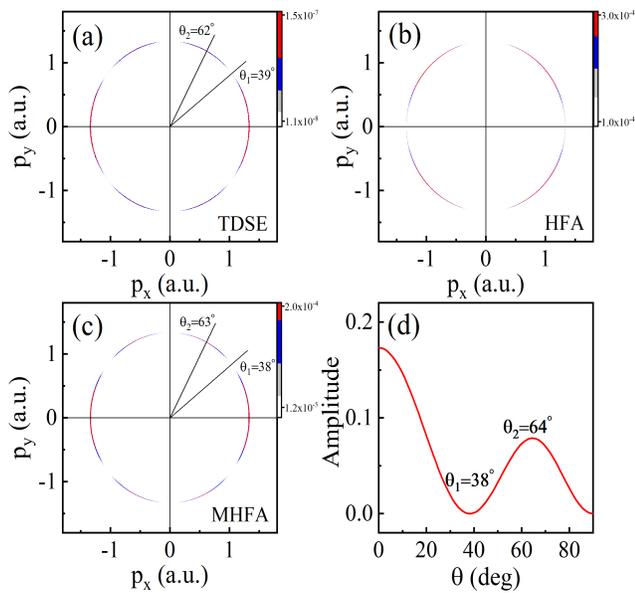}}}
\end{center}
\caption{PMDs of H$_2^+$ with $R=2$ a.u. and $\theta'=0^o$ obtained with different methods. (a): TDSE; (b): HFA; (c): MHFA. In (d), we also show the results of $M^2(\theta)$ of Eq. (11). The laser parameters used here are $I=1\times 10^{13}$ W/cm$^2$ and $\omega=2$ a.u.. The symbol $\theta_n$ with n being an integer denotes some extrema in quadrant 1 in the distributions.
}
\label{fig:g2}
\end{figure}

\begin{figure}[t]
\begin{center}
\rotatebox{0}{\resizebox *{8.5cm}{12cm} {\includegraphics {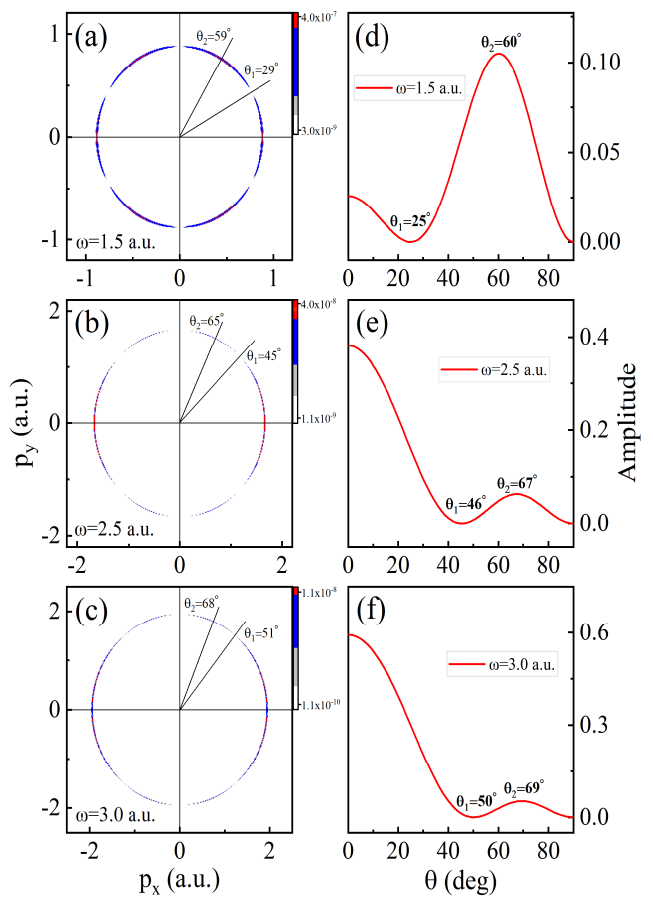}}}
\end{center}
\caption{PMDs of H$_2^+$ with $R=2$ a.u. and $\theta'=0^o$ obtained with different methods at different laser frequencies. The left column: TDSE; the right column: the corresponding results of $M^2(\theta)$ of Eq. (11). The laser intensity used here is $I=1\times 10^{13}$ W/cm$^2$ and the laser frequency $\omega$ used is as shown. The symbol $\theta_n$ with n being an integer denotes some extrema in quadrant 1 in the distributions.}
\label{fig:g3}
\end{figure}

With linear combination of atomic orbitals-molecular orbitals (LCAO-MO) approximation \cite{chen2009}, the dipole of H$_2^+$ in Eq. (2) can be written as
\begin{equation}\label{eq1}
\begin{split}
\textbf{e}\cdot \textbf{d}_i(\textbf{v})&=\cos (\textbf{v}\cdot \frac{\textbf{R}}{2})({\textbf{v}}\cdot \textbf{e})\frac{32\pi\kappa}{(\kappa^2+\textbf{v}^2)^3} \\ &=\cos (\frac{vR}{2}\cos\theta'')\cos\theta\frac{32\pi\kappa v}{(\kappa^2+v^2)^3}.
\end{split}
\end{equation}
Here, $\kappa=\sqrt{2I_p}$, $v=|\textbf{v}|$, $\theta''=\theta-\theta'$ and the term $\theta$ is the angle between the momentum $\textbf{v}$ and the vector $\textbf{e}$.

As discussed in \cite{chen2009}, due to the Coulomb effect,  the continuum-state wave function of $H_0$ near the position of the nucleus differs remarkably from the description of the plane wave. Because the integral of $\textbf{d}_i$ is mainly associated with the near-nucleus component of the continuum wave function, we use the effective momentum $\textbf{v}_k$ to replace $\textbf{v}$ in the interference term $\cos (\textbf{v}\cdot \frac{\textbf{R}}{2})$. This interference term describes the interference of the photoelectron with a momentum between these two atomic centers of the molecule.
Then we have
\begin{equation}\label{eq4}
\textbf{e}\cdot \textbf{d}_i(\textbf{v})=\cos (\frac{v_kR}{2}\cos\theta'')\cos\theta\frac{32\pi\kappa v}{(\kappa^2+v^2)^3}.
\end{equation}
Here,
\begin{equation}
\textbf{v}^2_k/2=\textbf{v}^2/2+I_p;  \textbf{v}_k/{v}_k=\textbf{v}/{v}.
\end{equation}
Clearly, in Eq. (9), the term
\begin{equation}\label{eq4}
M(\theta)=\cos (\frac{v_kR}{2}\cos\theta'')\cos\theta
\end{equation}
is only the component that depends on the emission angle $\theta$ and the alignment angle $\theta'=\theta-\theta''$.  This term $\cos (\frac{v_kR}{2}\cos\theta'')$ indicates that the interference occurs in terms of the Coulomb-related effective momentum $v_k$ instead of the Coulomb-free drift momentum $v$ observed in the detector. In the following, we will show that Eq. (11)
gives an applicable description of two-center interference in single-photon ionization of H$_2^+$ when the photon energy is not far larger than the
ionization potential $I_p$. In the following discussions, for simplicity, we call the predictions of Eq. (7) with Eq. (8) high-frequency approximation (HFA) and the predictions of  Eq. (7) with Eq. (9) modified high-frequency approximation (MHFA). It should be mentioned that for $\omega\gg I_p$ with $v_k\rightarrow v$, the predictions of MHFA become near to the HFA ones.

\begin{figure}[t]
\begin{center}
\rotatebox{0}{\resizebox *{8.5cm}{12cm} {\includegraphics {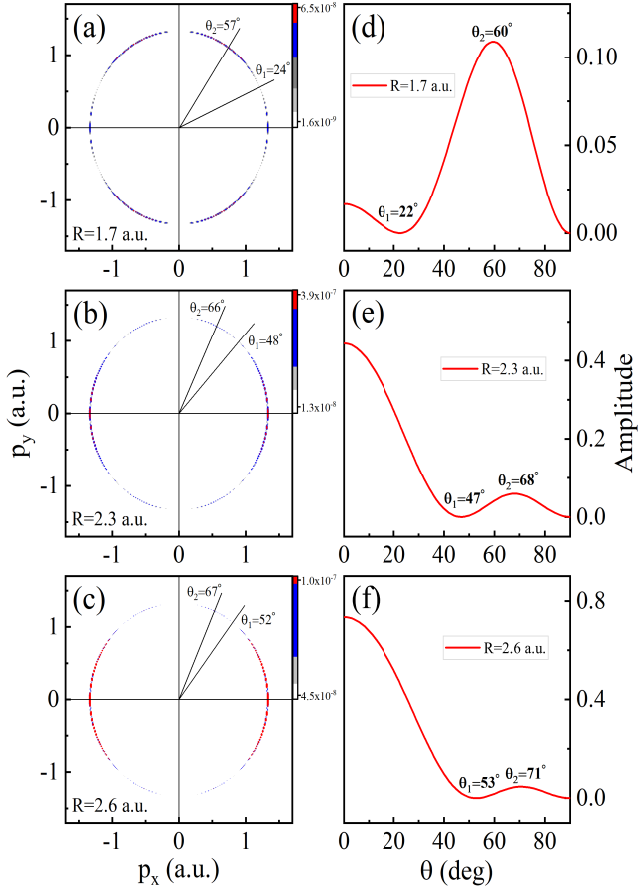}}}
\end{center}
\caption{PMDs of H$_2^+$ with $\theta'=0^o$ obtained with different methods at different internuclear distances $R$. The left column: TDSE; the right column: the corresponding results of $M^2(\theta)$ of Eq. (11). The laser parameters used here are $I=1\times 10^{13}$ W/cm$^2$ and $\omega=2$ a.u..
The internuclear distance $R$ used is as shown. The symbol $\theta_n$ with n being an integer denotes some extrema in quadrant 1 in the distributions.
}
\label{fig:g4}
\end{figure}
\begin{figure}[t]
\begin{center}
\rotatebox{0}{\resizebox *{8.5cm}{12cm} {\includegraphics {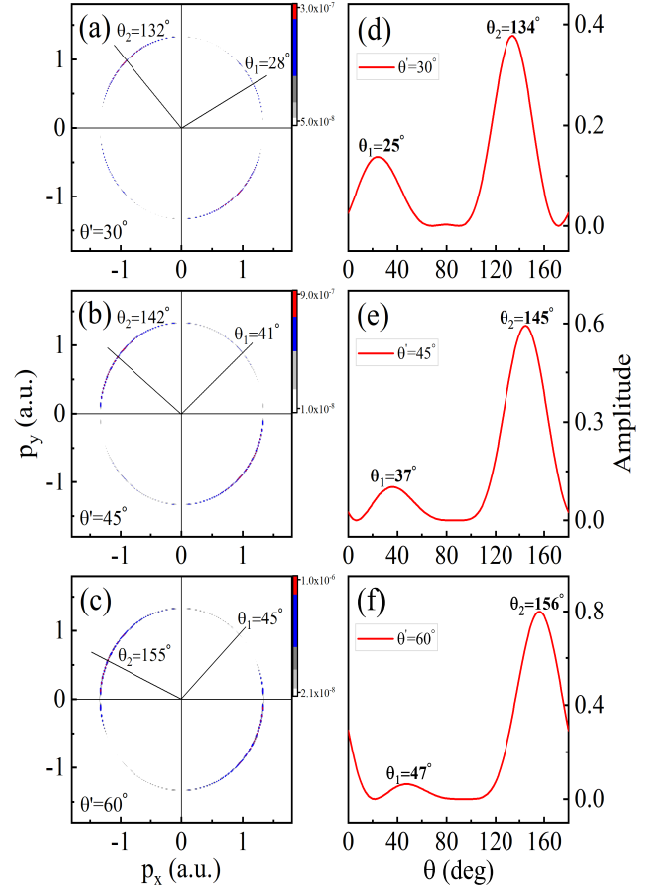}}}
\end{center}
\caption{PMDs of H$_2^+$ with $R=2$ a.u. obtained with different methods at different alignment angles $\theta'$. The left column: TDSE; the right column: the corresponding results of $M^2(\theta)$ of Eq. (11). The laser parameters used here are $I=1\times 10^{13}$ W/cm$^2$ and $\omega=2$ a.u..
The alignment angle $\theta'$ used is as shown. The symbol $\theta_n$ with n being an integer denotes some extrema in the upper half plane in the distributions.
}
\label{fig:g5}
\end{figure}

\section{Results and discussions}
In Fig. 2, we show TDSE results for PMDs of H$_2^+$ with comparing to the model predictions of HFA and MHFA.
For clarity, the $\log_{10}$ scale is used in plotting the calculated PMDs of $|c(\textbf{p})|^2$ in this paper.
The TDSE prediction in Fig. 2(a) shows a ring structure with some clear interference patterns.
The PMD of TDSE shows the maximal amplitude in the first quadrant for the momentum with the emission angle $\theta=0^o$.
The value of the momentum is $p_x=1.34$ a.u. with $p_y=0$ a.u. which agrees with the energy-conservation relation of $\textbf{p}^2/2=\omega-I_p$.
The TDSE distribution in quadrant 1 also shows a local minimum for the momentum with  the emission angle $\theta=\theta_1=39^o$ and  a local maximum for  $\theta=\theta_2=62^o$.
By comparison, the prediction of HFA  (i.e., Eq. (7) with Eq. (8) of the plane-wave momentum) in Fig. 2(b) does not show the maximal amplitude for $\theta=0^o$ and the local minimum at $\theta=39^o$ in quadrant 1. When the Coulomb-modified effective momentum is considered, the model results of MHFA in Fig. 2(c) agree with the TDSE ones with showing the maximal amplitude for $\theta=0^o$, a local minimum at $\theta=38^o$ and a local maximum at $\theta=63^o$ in quadrant 1. The results of  $M^2(\theta)$ of Eq. (11) in Fig. 2(d) are similar to those in Fig. 2(c). 
According to the above analyses and comparisons, 
the term $M(\theta)$ holds the essential interference-related component in Eq. (9).
In the following comparisons to TDSE for different laser and molecular parameters in Fig. 3 to Fig. 5,
for simplicity, we will show only the predictions of  $M^2(\theta)$.

In Fig. 3, we show PMDs of H$_2^+$ at different laser frequencies $\omega$. One can observe from the TDSE results (the left column of Fig. 3),
when the laser frequency changes, the interference patterns change remarkably.
For example, for $\omega=1.5$ a.u. (which is near to the ionization potential $I_p=1.1$ a.u.) in Fig. 3(a),
the maximal amplitude of the distribution  in quadrant 1 appears at $\theta=59^o$, a local minimum at $\theta=29^o$ and another
local maximum at $\theta=0^o$. By contrast, for $\omega=2.5$ a.u. in Fig. 3(b), the maximal amplitude of the distribution in quadrant 1 appears at $\theta=0^o$, the local minimum shifts to $\theta=45^o$ with another local maximum appearing at $\theta=65^o$. Similar phenomena to Fig. 3(b) are also seen in Fig. 3(c) of $\omega=3$ a.u.. These results are in good agreement with the predictions of $M^2(\theta)$ shown in the right column of Fig. 3.

We further change the internuclear distance $R$ in our simulations and relevant results are shown in Fig. 4.
The interference patterns in PMDs are also sensitive to $R$, as seen from the TDSE predictions in the left column of Fig. 4.
For example, for $R=1.7$ a.u. in Fig. 4(a), the maximal amplitude of the distribution in the first quadrant appears at $\theta=57^o$,
a local minimum at $\theta=24^o$ and a local maximum at  $\theta=0^o$. When the internuclear distance increases to $R=2.3$ a.u., as seen in Fig. 4(b), the maximal amplitude of the distribution in quadrant 1 shifts to $\theta=0^o$ and the local minimum shifts to $\theta=48^o$.
The interference patterns in PMDs for the case of $R=2.6$ a.u. in Fig. 4(c) are somewhat similar to those in Fig. 4(b).
These interference phenomena are also well reproduced by the results of  $M^2(\theta)$, as shown in the right column of Fig. 4.

Interference patterns of PMDs obtained at different alignment angles $\theta'$ are presented in Fig. 5.
These patterns with $\theta'=30^o, 45^o, 60^o$ differ remarkably from those of $\theta'=0^o$ in Fig. 1 to Fig. 4.
The TDSE distributions in the upper half plane  show the maximal amplitude around $\theta=132^o,142^o,155^o$, respectively,
and another local maximum around $\theta=28^o,41^o,45^o$, respectively, as seen in Fig. 5(a) to Fig. 5(c).
These phenomena are also in agreement with the predictions of  $M^2(\theta)$ in the right column of Fig. 5.

We have also extended our simulations to other laser intensities such as $I=5\times 10^{12}$ W/cm$^2$ and $I=5\times 10^{13}$ W/cm$^2$.
Results obtained are similar to those presented above, so we do not show them here. These extended simulations show that for the present cases of
single-photon ionization, the effect of two-center interference is not sensitive to the laser intensity.

\section{Conclusion}
In conclusion, we have studied single-photon ionization of aligned H$_2^+$ in high-frequency and low-intensity laser fields with different laser and molecular parameters. We have focused on the cases where the laser frequency is not far larger than the ionization potential of the molecule.
The PMDs of H$_2^+$ show clear interference patterns which change remarkably when the laser frequency, the internuclear distance as well as the
alignment angle change.
By developing a theory model applicable for high-frequency laser field, we show that the positions of interference minima and maxima in PMDs can not be well described by the interference
of the electronic wave with the drift momentum between these two atomic centers of the molecule. Instead, they can be quantitatively described
by the interference in terms of the effective momentum which considers the influence of the Coulomb potential on the continuum wave function near
the nuclei of the molecule.

These results indicate that the Coulomb potential plays an important role in the molecular-level two-center interference
in single-photon ionization when the photon energy is near to or somewhat lager than the ionization potential of the target molecule. 
One can not only retrieve the wave-function information of the ground state but also the continuum
state of the molecule from the interference patterns in PMDs.

As the interference of the emitting electronic wave between the two atomic centers occurs around the nuclei while the photoelectron momentum is probed in experiments far away from the nuclei, the interference effect also provides a chance for studying attosecond-resolved ionization dynamics of the electron inside a molecule in single-photon ionization.

It is interesting for studying single-photon ionization of an oriented polar molecule with a large permanent dipole in low-intensity high-frequency laser fields.
In addition, time-resolved studies on single-photon ionization of atoms and molecules using the non-perturbation approach developed in the paper are in progress.

\section*{Acknowledgements}
This work was supported by the National Natural Science Foundation of China (Grant Nos. 11904072, 12174239),
and the Fundamental Research Funds for the Central Universities of China (Grant No. 2021TS089).

\end{document}